\documentclass[12pt]{article}

\usepackage{epsfig,epstopdf,latexsym,amsfonts,amsmath,amsthm,amssymb,amsbsy,multirow,slashed,wasysym,textcomp,wrapfig,graphicx,psfrag,booktabs,bbm,comment}

\usepackage[toc]{appendix}

\usepackage{color}
\usepackage{datetime}
\usepackage[
      colorlinks=false,
      linkcolor=darkblue,  
      urlcolor=blue,    
      filecolor=blue,     
      citecolor=red,
linktocpage=true,
      pdfstartview=FitV,
      bookmarksopen=true    
      ]{hyperref}

\ifpdf
\fi

\numberwithin{equation}{section}

\def\coeff#1#2{\relax{\textstyle {#1 \over #2}}\displaystyle}

\def\IC{\mathbb{C}}

\def\IR{\mathbb{R}}
\def\ZZ{\mathbb{Z}}

\def\cA{{\cal A}}
\def\cB{{\cal B}}

\def\Neql#1{{\cal N}\!=\!{#1}}

\definecolor{cardinal}{rgb}{0.6,0,0}
\definecolor{darkgreen}{rgb}{0,0.5,0}
\definecolor{golden}{rgb}{0.92, 0.7, 0}
\definecolor{midnight}{rgb}{0, 0, 0.5}
\definecolor{darkblue}{rgb}{0.2, 0, 0.8}

\begin{document}

\begin{titlepage}

 \begin{flushright}
IPhT-T14/063\\
DCPT-14/11
\end{flushright}

\bigskip
\bigskip
\bigskip

\centerline{\Large \bf Coiffured Black Rings}
\smallskip

\medskip
\bigskip
\bigskip
\centerline{{\bf Iosif Bena$^1$,   Simon F. Ross$^2$ and Nicholas P. Warner$^{1,3,4}$}}
\centerline{{\bf }}
\bigskip
\centerline{$^1$ Institut de Physique Th\'eorique, }
\centerline{CEA Saclay, CNRS-URA 2306, 91191 Gif sur Yvette, France}
\bigskip
\centerline{$^2$ Centre for Particle Theory, Department of Mathematical Sciences }
\centerline{Durham University, South Road, Durham DH1 3LE, UK}
\bigskip
\centerline{$^3$ Department of Physics and Astronomy}
\centerline{University of Southern California} \centerline{Los
Angeles, CA 90089, USA}
\bigskip
\centerline{$^4$ Institut des Hautes Etudes Scientifiques}
\centerline{91440 Bures-sur-Yvette, France}
\bigskip
\bigskip
\centerline{{\rm iosif.bena@cea.fr,~s.f.ross@durham.ac.uk,~warner@usc.edu} }
\bigskip
\bigskip

\begin{abstract}
\noindent We describe a new type of hair on supersymmetric black string and black ring solutions, which produces the largest known violation of black hole uniqueness, parameterized by an arbitrary function and hence an infinite number of continuous parameters. The new solutions can have  non-trivial density profiles for the electric fields along the horizon, and yet have a geometry that is regular, although generically not infinitely differentiable, at the horizon.  Both neutral and charged probes can cross the horizon without experiencing divergent forces. We also find restricted examples, parameterized by a few arbitrary continuous parameters, where the charge densities fluctuate but the metric does not and hence is completely differentiable. Our new class of solutions owes its existence to a mechanism reminiscent of the Q-ball: in the simplest examples the metric has more symmetry than the matter that supports it.
\end{abstract}

\end{titlepage}

\section{Introduction}

Five and six-dimensional supergravity theories have a surprisingly rich solution space. The construction of smooth, horizonless microstate geometries in these theories has led to a new perspective on the microstates of black holes. The simplest class of microstate geometries are supertubes \cite{Mateos:2001qs}, in which two types of branes form a bound state that has a non-trivial dipole profile of arbitrary shape. In the D1-D5 frame this dipole charge corresponds to a Kaluza-Klein monopole, and the resulting smooth geometry is parameterized by several continuous functions that depend on the embedding of the world-volume of this monopole in the spacetime \cite{Lunin:2001fv,Lunin:2002iz}. One can imagine superposing several different types of supertubes to obtain a black ring \cite{Bena:2004de,Elvang:2004ds,Gauntlett:2004qy}, but if one tries to do this with supertubes that have a non-trivial charge profile and obtain a BPS black ring with varying charge densities \cite{Bena:2004td} one finds that the horizon becomes singular \cite{Horowitz:2004je}. Thus, it appears that the horizon of a black ring cannot support the kind of non-trivial charge hair that supertubes can possess. 

As a consequence of this, for almost ten years the prevailing lore has been that, while the space of black objects in five-dimensional supergravity is much richer than in four dimensions, the violations of black-hole uniqueness in five dimensions come only from the {\it discrete} choices of the black-ring dipole charges and that, not only are there  no infinite-dimensional moduli spaces worth of hair (as one might have expected from supertubes), but there is not even a finite-dimensional moduli space of solutions parameterized by a finite number of {\it continuous} parameters.  In this paper we show that this lore this is incorrect by explicitly constructing some very hairy black-ring solution that can be  {\it coiffured}  so as to preserve a lot of the hair and yet have well-behaved horizons.   In particular, we  find families of solutions that depend upon  several continuous parameters for which the metric is $C^{\infty}$ across the horizon.  More generally, we argue that the hair can be arranged so that  the Riemann tensor remains at least continuous across the horizon while the hair still depends upon an arbitrary continuous function.

The key idea behind this construction comes from our recent discovery of a new BPS object in M-theory - the {\em magnetube} \cite{Bena:2013ora}. This object has M5 and momentum charges (which are magnetic in five dimensions), as well as several electric M2 charge densities that can oscillate between positive and negative values along the M5-P common direction. The M2 charge densities can have non-trivial oscillating profiles along this direction but these density profiles can be arranged so that the total energy-momentum tensor they source does not oscillate and yet the overall density configuration has an arbitrary function's-worth of freedom.

In this paper, we consider putting similar oscillating electric charges on supersymmetric black strings and black rings. We will show that, unlike the simple single-species fluctuations considered in  \cite{Bena:2004td,Horowitz:2004je}, the profiles of the oscillating species can be chosen such that the first few terms in the near-horizon expansion of the metric are independent of the angle around the ring. Furthermore, in the simplest example, the metric can be made completely independent of this angle, even though  the Maxwell fields fluctuate. Thus these solutions are rather reminiscent of Q-Balls  \cite{Coleman:1985ki,Friedberg:1986tq} and hence we dub them ``O-rings.'' 

The smoothness of the metric at the O-ring horizon is achieved by arranging that the energy-momentum tensor of the fields is either independent of, or has much weaker dependence upon,  the direction in which the other fields fluctuate.  The curvatures at the horizon then remain finite, and an infalling observer can survive the passage across the horizon. The electric fields will oscillate infinitely many times along the trajectory of an infalling observer, but since the fields remain finite, the force on a charged particle falling across the horizon is also finite, and given the oscillating nature of the fields the integrated force is also finite. Thus by ``coiffing'' a suitably rich variety of hair, in the form of charge densities, one can arrange for regularity of the gravitational background. 

Our very hairy black rings provide by far the largest known violation of black-hole uniqueness, being regular black objects with an arbitrary continuous function's worth of hair. While our solutions are  supersymmetric, we do not anticipate obstructions in adapting and generalizing our results to non-supersymmetric black rings.  It would be very interesting to investigate this in detail and  determine the possible fluctuation directions that can exploit the mechanism underpinning O-rings.

In Section 2 we review the basic supergravity setting for our new class of solutions. Since horizon regularity is a ``near-ring'' issue we start by examining the simplest possible solutions: black strings in five dimensions.  In Section 3  we review the non-oscillating strings (that have a $\IR^3 \times S^1$ base space) and describe, in some detail, the horizon regularity and the adapted coordinates allowing continuation across the horizon.  In Section 4 we find the simplest BPS oscillating strings: O-rings in $\IR^3 \times S^1$ and derive the constraints on the charge densities for  smoothness across the horizon.  Section 5 contains the solutions for O-rings in $\IR^4$: there are no additional issues of smoothness compared to the O-rings in  $\IR^3 \times S^1$.  In Section 6 we consider more general classes of O-rings in $\IR^4$ and perform the computations in the more traditional separable black-ring coordinate system.  The impatient reader who is familiar with the standard description of black rings and does not care about their construction as BPS solutions with a Gibbons-Hawking space, could skip directly to this section as it is relatively self-contained.  Section 7 contains some concluding comments.

\section{The supergravity setting of the new BPS solutions}

\subsection{The supergravity action}

As in \cite{Bena:2013ora}, we use  $\mathcal N = 2$ supergravity in five dimensions coupled to vector multiplets, except that here we consider more vector multiplets.  We will thus follow the conventions of \cite{Gutowski:2004yv,Gauntlett:2004qy}. The action of $\Neql2$, five-dimensional supergravity coupled to $N$ $U(1)$ gauge fields is
\begin{eqnarray}
  S = \frac {1}{ 2 \kappa_{5}} \int\!\sqrt{-g}\,d^5x \Big( R  -\coeff{1}{2} Q_{IJ} F_{\mu \nu}^I   F^{J \mu \nu} - Q_{IJ} \partial_\mu X^I  \partial^\mu X^J -\coeff {1}{24} C_{IJK} F^I_{ \mu \nu} F^J_{\rho\sigma} A^K_{\lambda} \bar\epsilon^{\mu\nu\rho\sigma\lambda}\Big) \,,
  \label{5daction}
\end{eqnarray}
with $I, J =0, \dots, N$.  The extra photon lies in the gravity multiplet and so there are only $N$  independent scalars. It is, however, convenient to parametrize them by $N+1$ scalars $X^I$,  satisfying  the  constraint
\begin{equation}
\coeff{1}{6} \, C_{IJK} X^I \,  X^J \, X^K  ~=~  1 \,.
\label{Xconstr}
\end{equation}
Following standard practice, introduce
\begin{equation}
X_I ~\equiv~  \coeff{1}{6} \, C_{IJK}   X^J \, X^K \,.
\label{Xdown}
\end{equation}
The  scalar kinetic term can then be written as
\begin{equation}
  Q_{IJ} ~=~  \coeff{9}{2} \, X_I \, X_J ~-~ \coeff{1}{2} \, C_{IJK} X^K \,.
\label{scalarkinterm}
\end{equation}
The Chern-Simons structure constants are required to satisfy the constraint
\begin{equation}
 C_{IJK} \,   C_{J' (LM} \,  C_{PQ)K'}  \, \delta^{J J'} \,  \delta^{K K'} ~=~  \coeff{4}{3} \, \delta_{I(L}\, C_{MPQ)}    \,.
\label{Cconstr}
\end{equation}
It is also convenient to define
\begin{equation}
 C^{IJK}~=~   \delta^{I I'} \,  \delta^{J J'} \,  \delta^{K K'}  \, C_{I'J'K'}    \,.
\label{Cupper}
\end{equation}
Using the constraint (\ref{Cconstr}), one can  show that the inverse, $Q^{IJ}$, of $Q_{IJ}$ is given by:
\begin{equation}
Q^{IJ} ~=~ 2 \, X^I \, X^J ~-~ 6\, C^{IJK} X_K    \,, 
\label{Qinv}
\end{equation}
and one can show that
\begin{equation}
\coeff{1}{6} \, C^{IJK} X_I \,  X_J \, X_K  ~=~  \frac{1}{27} \,.   
\label{Cconstr2}
\end{equation}

As in \cite{Giusto:2012gt,Vasilakis:2012zg, Bena:2013ora}, we will focus on theories that can be obtained from eleven-dimensional supergravity reduced on a $T^6$.  The Maxwell fields then descend from the tensor gauge field, $C^{(3)}$, via harmonic $2$-forms on $T^6$ and the  structure constants $C_{IJK}$ are  given by the   intersection product of the  homology cycles and the $X^I$ are moduli of the $T^6$.  The restriction to torus compactifications is made for simplicity and convenience and it will be evident that rather richer classes of O-ring solutions can  be obtained from Calabi-Yau compactifications with more complicated cohomology and intersection forms.  
  
\subsection{The supersymmetry conditions}

We start with the most general stationary five-dimensional metric:
\begin{equation}
ds_5^2 ~=~ -Z^{-2} \,(dt + k)^2 ~+~ Z \, ds_4^2  \,,
\label{metAnsatz}
\end{equation}
where $Z$ is simply a convenient warp factor.   Supersymmetry implies that the metric $ds_4^2$ on the spatial base manifold, $\cB$, must be hyper-K\"ahler.

One now defines $N+1$ independent functions, $Z_I$, by
\begin{equation}
Z_I ~=~  3 \, Z \, X_I  \,,
\label{ZIdefn}
\end{equation}
and then (\ref{Cconstr2}) implies
\begin{equation}
Z   ~=~  \big(\coeff{1}{6} \, C^{IJK} Z_I \, Z_J \, Z_K\big)^{\frac{1}{3}}\,.   
\label{Zreln}
\end{equation}
It is more convenient to think of the solution as parametrized by the $N+1$ independent scalars, $Z_I$; the warp factor is then determined by (\ref{Zreln}).

Supersymmetry requires that the Maxwell potentials all have the form
\begin{equation}
A^{(I)}   ~=~  - \coeff{1}{2} \, Z^{-3}\, C^{IJK} \, Z_J \, Z_K (dt +k)  ~+~ B^{(I)}  \,,
\label{AAnsatz}
\end{equation}
where $B^{(I)}$ are purely magnetic components on the spatial base manifold, $\cB$.  One defines the magnetic field strengths accordingly: 
\begin{equation}
\Theta^{(I)}   ~=~ d \, B^{(I)}  \,.
\label{Thetadefn}
\end{equation}

Having made all these definitions, the BPS equations take their canonical linear form \cite{Bena:2004de}:
\begin{eqnarray}
 \Theta^{(I)}  &=&  \star_4 \, \Theta^{(I)} \label{BPSeqn:1} \,, \\
\nabla^2_{(4)}  Z_I &=&  \coeff{1}{2} \,C_{IJK} \, \star_4 \Theta^{(J)} \wedge
\Theta^{(K)} \label{BPSeqn:2} \,, \\
 dk ~+~  \star_4 dk &=&  Z_I \,  \Theta^{(I)}\,,
\label{BPSeqn:3}
\end{eqnarray}
where $\star_4$ is the Hodge dual in the four-dimensional base metric $ds_4^2$, and  $\nabla^2_{(4)}$ is the (four-dimensional) Laplacian in this metric.

In Section \ref{CanonCoords}  we will take $\cB$ to be flat $\IR^4$ written in the canonical black-ring coordinates. However, it is very useful for our initial analysis to take the metric on $\cB$ to be a Gibbons-Hawking  metric because this will enable us to  move easily between the solutions on  $\mathbb R^3 \times S^1$ and $\mathbb R^4$.  It will  also make subsequent generalizations of our analysis to black rings in Taub-NUT \cite{Elvang:2005sa,Gaiotto:2005xt,Bena:2005ni} and to multi-centered solutions relatively straightforward.

In the next sections, the four-dimensional metric will therefore be of the form: 
\begin{equation}
ds_4^2 = V^{-1} \, (d\psi + \cA)^2 +  V\, d\vec y \cdot d\vec y   = V^{-1} (d\psi + \cA)^2 + V (dr^2 + r^2 (d\theta^2 + \sin^2 \theta d\phi^2)) \,,
\label{GHmetric}
\end{equation}
with 
\begin{equation}
\vec \nabla \times \vec \cA ~=~ \vec \nabla V   \,.
\label{AVreln}
\end{equation}

The magnetic potentials that solve the supersymmetry conditions are: 
\begin{equation}
B^I = \frac{K^I}{V} (d \psi+ \cA) + \vec \xi^I\cdot d \vec{y} + \alpha^I 
\end{equation}
with $\nabla_3^2 K^I = 0$, $\vec \nabla \times \vec \xi^I = - \vec \nabla K^I$, and $d \alpha^I = 0$. The closed form, $\alpha^I$, is usually set to zero, but we will see later that we need to take it to be non-zero in order to have regular gauge potentials at the horizon. The functions $Z_I$ are
\begin{equation}
Z_I = \frac{1}{2V}  C_{IJK} K^J K^K + L_I 
\end{equation}
with $\nabla_4^2 L_I = 0$. We will assume that $K^I$ and $V$ are independent of $\psi$, and depend only on the coordinates $\vec{y}$ on the base. In a suitable gauge, the last supersymmetry equation is satisfied by taking  $k = \mu d \psi + \omega$, with 
\begin{equation}
\mu = \frac{1}{6 V^2} C_{IJK} K^I K^J K^K + \frac{1}{2 V} K^I L_I + M
\end{equation}
and 
\begin{equation}
\vec {\mathcal{D}} \times \vec \omega ~+~ V \partial_\psi \vec \omega ~=~ V \vec {\mathcal{D}} M - M\vec {\mathcal{D}} V +\frac{1}{2} \, \sum_{I=1}^{N+1} \big( K^{I} \vec {\mathcal{D}} L_{I} - L_{I}   \vec {\mathcal{D}} K^{I} \big) \,.
\label{simpomegaeqn}
\end{equation}
where
\begin{equation}
\vec {\mathcal{D}} ~\equiv~ \vec \nabla ~-~   \vec \cA \,\partial_\psi   \, ,
\label{simpcovD}
\end{equation}
and
\begin{equation}
\nabla^2_{(4)}    M ~=~ 0 \,.
\label{harmonicLM}
\end{equation}
%

\subsection{Adding species on the $T^4$}

From the eleven-dimensional perspective, we are going to add extra Maxwell fields coming from three-form potentials with two legs on the $T^4$ defined by $(x^5,x^6,x^7,x^8)$ but leave the fields on the other $T^2$, defined by $(x^9,x^{10})$  unchanged and supporting only one Maxwell field, which we have labelled as $A^{(0)}$.   Thus the only non-zero components of the intersection product $C_{IJK}$ are   
\begin{equation}
C_{0JK}  ~=~  \widehat C_{JK}  ~=~  \widehat C_{KJ} \,,
\label{RecC}
\end{equation}
and one has \cite{Bena:2013ora}:
\begin{equation}
\widehat C_{IJ}  ~=~  
\begin{pmatrix} 
0&1&0&0\\  1&0&0&0\\ 0&0&-1&0\\ 0&0&0&-1
\end{pmatrix}   \,,
\label{hatCform}
\end{equation}
which satisfies $\widehat C_{IJ}\widehat C_{KL} \delta^{JK} = \delta_{IL}$.  
Observe that (\ref{Zreln}) now implies that the space-time metric warp factor, $Z$, is given by
\begin{equation}
Z^3   ~=~  \coeff{1}{2} \, Z_0 \, \big( \widehat C^{IJ}   Z_I \, Z_J \big)~=~   Z_0 \, \big(   Z_1 \, Z_2  - \coeff{1}{2} \, (Z_3^2  + Z_4^2) \big)\,.   
\label{Zreln2}
\end{equation}
It is also convenient to define the quadratic combination
\begin{equation}
P   ~\equiv~ \widehat C^{IJ}   Z_I \, Z_J~=~  \big(    Z_1 \, Z_2  - \coeff{1}{2} \, (Z_3^2  + Z_4^2) \big)\,.   
\label{Pdefn}
\end{equation}

Since we are considering black objects, we will work entirely in the five-dimensional theory; we do not need to uplift to the higher-dimensional description.

\section{String with no oscillation}

We will first discuss the simplest solutions: black strings with oscillating species, so that we may examine horizon smoothness without the algebra becoming unnecessarily complicated.  We therefore take the base metric to be simply a flat metric on $\IR^3 \times S^1$, which means we set $V \equiv 1$ and $\cA\equiv 0$ in the metric \eqref{GHmetric}. The string is then aligned along the $\psi$ direction. A useful starting point is to review the horizon smoothness of the non-oscillating electrically and magnetically charged black string solution \cite{Bena:2004wv}, which is independent of $\psi$ and is given by the harmonic functions:
\begin{equation}
K^I = \frac{p}{r}, \quad L_I = 1 + \frac{q}{r}
\end{equation}
for $I = 0,1,2$, and $K^I = 0$, $L_I = 0$ for $I=3,4$. The functions in the metric are then 
\begin{equation}
Z = 1 + \frac{q}{r} + \frac{p^2}{r^2} 
\end{equation}
and 
\begin{equation}
\mu = \frac{p^3}{r^3} + \frac{3pq}{2r^2} + \frac{3p}{r}, 
\end{equation}
choosing $M = \frac{3p}{2r}$ so that we can simply solve \eqref{simpomegaeqn} by taking $\omega=0$. The scalars are all constants, $X^I = 1$ for $I=0,1,2$, $X^I = 0$ for $I=3,4$. 

The metric and gauge fields are clearly finite for $r \neq 0$. The solution has a coordinate singularity at $r=0$, and we want to see  how this corresponds to an event horizon. As $r \to 0$, the metric is asymptotic to
\begin{equation}
ds^2  \approx -\frac{r^4}{p^4} dt^2  -\frac{2r}{p} dt d \psi + \frac{3}{4p^2}(q^2 - 4p^2) d\psi^2  + \frac{p^2}{r^2} dr^2 + p^2 (d\theta^2 + \sin^2 \theta d\phi^2) \,,
\end{equation} 
which is singular in the $dr$ direction and degenerate in the $dt$ direction.   To see that this is simply a coordinate singularity one can use a simple generalization of the 
Eddington-Finkelstein  coordinate system:  
\begin{equation} \label{ct}
t = v + \frac{a}{r} + b \log(r/p), \quad \psi = \hat \psi + c \log(r/p). 
\end{equation}
Expand the metric into power series in $r$: 
\begin{eqnarray}
g_{t\psi} &=& r (g_{t\psi}^{(0)} + r g_{t\psi}^{(1)} + r^2 g_{t\psi}^{(2)})\,, \\
g_{\psi\psi} &=& g_{\psi\psi}^{(0)} + r g_{\psi\psi}^{(1)} + r^2 g_{\psi\psi}^{(2)}\,, \nonumber \\
g_{rr} &=& \frac{1}{r^2} (g_{rr}^{(0)} + r g_{rr}^{(1)} + r^2 g_{rr}^{(2)})\,. \nonumber
\end{eqnarray}
To the first few orders in $r$, the components of interest in the new coordinate system are 
\begin{eqnarray} \label{rpsi}
g'_{r\hat \psi} &=& -2 \left( \frac{a}{r^2} - \frac{b}{r} \right) g_{t\psi} + 2 \frac{c}{r} g_{\psi \psi} \\ &=& \frac{1}{r} (2 c g_{\psi\psi}^{(0)} - 2 a g_{t\psi}^{(0)}) + (2 c g_{\psi\psi}^{(1)} - 2 a g_{t\psi}^{(1)} + 2 b g_{t\psi}^{(0)} ) + \ldots \nonumber 
\end{eqnarray}
and 
\begin{eqnarray} \label{rr}
g'_{rr} &=& \left( \frac{a}{r^2} - \frac{b}{r} \right)^2 g_{tt} - 2 \left( \frac{a}{r^2} - \frac{b}{r} \right) \frac{c}{r} g_{t\psi} + \frac{c^2}{r^2} g_{\psi\psi} + g_{rr} \\ &=& \frac{1}{r^2} (-2ac g_{t\psi}^{(0)} + c^2 g_{\psi\psi}^{(0)} + g_{rr}^{(0)} ) + \frac{1}{r} (- 2 ac g_{t\psi}^{(1)} + 2bc g_{t\psi}^{(0)} +c^2 g_{\psi\psi}^{(1)} + g_{rr}^{(1)}) \nonumber \\ &&+ (-a^2 p^{-4} -2 ac g_{t\psi}^{(2)} + 2bc g_{t\psi}^{(1)} + c^2 g_{\psi\psi}^{(2)} + g_{rr}^{(2)})  + \ldots \,. \nonumber 
\end{eqnarray}
There is a potential singularity in $g'_{r\hat \psi}$ at order $r^{-1}$ and cancelling this fixes $a$ in terms of $c$: 
\begin{equation} \label{aceq}
a ~=~ \frac{g_{\psi\psi}^{(0)}}{g_{t\psi}^{(0)}}\,  c  ~=~ - \frac{3 \, c}{4p} (q^2 - 4 p^2).
\end{equation}
Then there are potential singularities in $g'_{rr}$ at both order $r^{-2}$ and $r^{-1}$, which fix the other two parameters $c, b$, 
\begin{equation}
c   ~=~ \frac{2 p^2}{\sqrt{3} \sqrt{q^2 - 4p^2}}, \quad b ~=~ \frac{(q^3-3p^2q)}{\sqrt{3} p \sqrt{q^2-4p^2}}. 
\end{equation}
With these choices, the metric near $r=0$ reduces to
\begin{equation} \label{newmet}
ds^2 \approx -\frac{2c}{p} dv dr + f d\hat \psi^2 + 2 h dr d\hat \psi + j dr^2 + p^2 (d\theta^2 + \sin^2 \theta d\phi^2), 
\end{equation}
where 
\begin{equation} \label{fhk}
f = g_{\psi\psi}^{(0)}, \quad h = 2 c g_{\psi\psi}^{(1)} - 2 a g_{t\psi}^{(1)} + 2 b g_{t\psi}^{(0)} , \quad j = -a^2 p^{-4} -2 ac g_{t\psi}^{(2)} + 2bc g_{t\psi}^{(1)} + c^2 g_{\psi\psi}^{(2)} + g_{rr}^{(2)}.
\end{equation}
The determinant of the metric is $g \approx - f c^2 p^2 \sin^2 \theta$. The metric is thus completely regular and non-degenerate near $r=0$. 

We should also consider the regularity of the matter fields. For the vector fields, $A^3 = A^4 = 0$ while 
\begin{equation}
A^0 = A^1 = A^2 = A = Z^{-1} (dt + \mu d\psi) - \frac{p}{r} d\psi + p \cos \theta d\phi + \alpha \,.
\end{equation}
(Note that the vector field, $A$ here should not be confused with $\cA$ in (\ref{GHmetric}).)
All the components are finite at $r=0$ in the original coordinates, but we need to consider the behaviour in the regular coordinates:
\begin{equation} 
A = Z^{-1} \left ( dv - \frac{a}{r^2} dr - \frac{b}{r} dr \right) + \left( \mu Z^{-1} - \frac{p}{r} \right) \left( d \hat \psi + \frac{c}{r} dr \right)   + p \cos \theta d\phi + \alpha.
\end{equation}
Now $Z \sim r^{-2}$, so the first part is finite, but a constant term in $A_\psi$ could produce a divergence in $A_r$ in the new coordinates. Indeed $\mu Z^{-1} - \frac{p}{r} \approx \frac{q}{2p}$. But this can be cancelled by taking advantage of the freedom to introduce a closed form, setting $\alpha = - \frac{q}{2p} d\psi$.\footnote{Note this is not a gauge transformation, as $\psi$ is a periodic coordinate, so this addition is changing the holonomy around the circle. Thus the solutions that is regular at the horizon has some non-trivial holonomy at infinity.} Then one has: 
\begin{equation} 
A = Z^{-1} \left ( dv - \frac{a}{r^2} dr - \frac{b}{r} dr \right) + \left( \mu Z^{-1} - \frac{p}{r} - \frac{q}{2p}  \right) \left( d \hat \psi + \frac{c}{r} dr \right)  + p \cos \theta d\phi \,,
\end{equation}
which is finite at the horizon, $r=0$, in the new coordinates. Thus, this solution describes a regular supersymmetric black string with electric and magnetic charges.

\section{Oscillating black string}

We now add  oscillations to the black string solution, promoting the  $L_I$, which control the electric charges, to functions of  $\psi$.  We will not change the magnetic charge functions, so 
\begin{equation} \label{kfns}
K^0 = K^1 = K^2 = \frac{p}{r}, \quad K^3 = K^4 = 0 \,,
\end{equation}
but we now take
\begin{equation} \label{lfns} 
L_0 = 1 + \frac{q}{r}\,, \qquad L_{I} = 1 + \frac{q}{r} + \frac{F_I(\psi,r)}{r}
\end{equation}
for $I=1,2,3,4$. The $L_{I}$ are required to be solutions of $\nabla_4^2 L_{I} = (q+ \rho_I(\psi)) \delta^3(\vec x)$.   

Since we are working on a flat base, it is easy to write an explicit integral form for the solutions:
\begin{equation} \label{FG}
F_I(\psi,r) = r \int_{0}^{\Delta \psi} \frac{ \rho_I(\psi') d\psi'}{\pi (r^2 + (\psi - \psi')^2)}\,.
\end{equation}
It is useful to note that if $\rho_I(\psi) = \mathrm{Re} [ \rho_I(z)]$, where $\rho_I(z)$ has no poles in the upper-half plane, then the result of doing this integral can be written as $F_I(r, \psi) = \mathrm{Re} [\rho_I(\psi + i r)]$ \cite{Horowitz:2004je}.  We will subsequently assume that the charge distributions are purely oscillating, so the integral of $\rho_I$ over the circle vanishes; this part carries no net charge. This implies $\int d\psi F_I = 0$ as well. 

Our goal is to suppress as much of the $\psi$-dependence from the metric as possible and to that end we will choose the densities such that  the fluctuations cancel in the source term of (\ref{BPSeqn:3}) and thus the angular momentum vector is $\psi$-independent. This is  achieved by choosing the opposite oscillating densities: 
\begin{equation} \label{densconstr1}
 \rho_2(\psi)  ~=~  - \rho_1(\psi)   \,,
\end{equation}
which implies that $F_2 = - F_1$, exactly as in the magnetube solutions in \cite{Bena:2013ora}. 
 
The functions $Z_I$ are
\begin{equation}
Z_0 = 1 + \frac{q}{r} + \frac{p^2}{r^2}, \quad Z_{1,2} = 1 + \frac{q}{r}  \pm \frac{F_1}{r} + \frac{p^2}{r^2}, \quad Z_{3,4} =  \frac{F_{3,4}}{r}, 
\end{equation}
and
\begin{equation}
Z^3 = Z_0 (Z_1 Z_2 - \coeff{1}{2}( Z_3^2+  Z_4^2)) = \left( 1 + \frac{q}{r} + \frac{p^2}{r^2} \right) \left(  \left( 1 + \frac{q}{r} + \frac{p^2}{r^2} \right)^2 - \frac{(F_1^2 +\frac{1}{2}( F_3^2+ F_4^2))}{r^2} \right).
\end{equation}
We can again take 
\begin{equation}
\mu = \frac{p^3}{r^3} + \frac{3pq}{2r^2} + \frac{3p}{r}, 
\end{equation}
choosing $M= \frac{3p}{2r}$. The choice (\ref{densconstr1}) implies that there are no $\psi$-dependent sources in the equation  \eqref{simpomegaeqn} for $\omega$, so with this choice for $M$ this equation is again satisfied by taking $\omega = 0$.  Note that $\psi$-dependence in the metric only enters through $Z$. 

In fact, it is possible to find particular solutions with non-trivial $\psi$-dependence in the fields but in which there is no $\psi$-dependence at all in the metric. For example, if one takes
\begin{equation}
\rho_1(\psi) = - \rho_2(\psi) =  Q \cos (k \psi), \quad \rho_3  = \rho_4 = Q \sin (k \psi), 
\end{equation}
then 
\begin{equation}
F_1 = Q  \cos k \psi \, e^{-k r}, \quad F_3 = F_4 = Q  \sin k \psi\,  e^{-k r} \Rightarrow F_1^2 + \frac{1}{2}(F_3^2 + F_4^2) = Q^2 e^{-2kr}, 
\end{equation}
and $Z$ and hence the metric  will be completely independent of $\psi$. The non-trivial $\psi$-dependence would appear just in the gauge field and scalars.   Thus $\frac{\partial}{\partial \psi}$ is a Killing vector but is not a symmetry of the complete solution.  It is in this sense that our solutions are analogous to Q-balls  \cite{Coleman:1985ki,Friedberg:1986tq}.

More generally, if the source functions are some arbitrary periodic harmonic functions, then the $F^I$ will decay exponentially at large $r$ and will be non-singular near $r=0$, where $F^I(r, \psi) =  \mathrm{Re} [\rho_I(\psi + i r)] \approx \rho_I(\psi) + \mathcal{O}(r)$. We can therefore write the combination appearing in $Z$  in some power series expansion about $r=0$, 
\begin{equation}
F_1^2 + \coeff{1}{2} (F_3^2 + F_4^2) = F^{(0)} (\psi) + F^{(1)}(\psi) \,  r + F^{(2)}(\psi) \, r^2 + \ldots
\end{equation}
where $F^{(0)}(\psi) = \rho_1^2 + \frac{1}{2} (\rho_3^2 + \rho_4^2)$. 

\subsection{Regularity}

\subsubsection{The scalars}

The scalars are given by:
\begin{equation} \label{scalars}
X^0 = \frac{(Z_1 Z_2 - \frac{1}{2}(Z_3^2 + Z_4^2))^{1/3}}{Z_0^{2/3}}, \quad X^1 = \frac{Z_0 Z_2}{Z^2}, \quad X^2 = \frac{Z_0 Z_1}{Z^2}, \quad X^3 = \frac{Z_0 Z_3}{Z^2}, \quad X^4= \frac{Z_0 Z_4}{Z^2}.
\end{equation}
Since $Z_0$ is unchanged, $X^0$ clearly remains finite everywhere and, assuming $F_1^2 + \frac{1}{2}( F_3^2 + F_4^2)$ is exponentially decaying, we can choose $q$ big enough so that $Z$ is finite everywhere so all the scalars are regular for $r \neq 0$. As we approach $r=0$, $X^I \to 1$ for $I=0,1,2$ and $X^I \to 0$ for $I=3,4$. 

\subsubsection{The metric}

As before, the metric has an apparent singularity and degeneracy at $r=0$. As $r \to 0$, we have:
\begin{equation}
ds^2  \approx -\frac{r^4}{p^4} dt^2  -\frac{2r}{p} dt d \psi + \frac{1}{4 p^2} (3 q^2 - 12 p^2 - 4 F^{(0)})  d\psi^2  + \frac{p^2}{r^2} dr^2 + p^2 (d\theta^2 + \sin^2 \theta d\phi^2).   
\end{equation} 
One wants to make the same coordinate transformation \eqref{ct} to attempt to make the metric regular at $r=0$. It was shown in \cite{Horowitz:2004je} that in the previous attempt to build hairy black rings \cite{Bena:2004de,Bena:2004td}  this coordinate transformation was obstructed, as the leading term in $g_{\psi\psi}$ was $\psi$ dependent, so we can't set $a = c g_{\psi\psi}^{(0)}/g_{t\psi}^{(0)}$ as in \eqref{aceq} for constant $a, c$. Similarly here there is generically $\psi$ dependence in $g_{\psi\psi}$; but we can choose charge densities such that  the sum of the squares $F^{(0)}(\psi) = \rho_1^2 + \frac{1}{2} (\rho_3^2 + \rho_4^2)$ is a constant. This condition of constant amplitude for the fluctuating densities is familiar from our study of magnetubes \cite{Bena:2013ora}. 

Imposing this constraint, we can eliminate the potential singularity in $g'_{r\psi}$ in \eqref{rpsi} by setting 
\begin{equation}
a =  \frac{g_{\psi\psi}^{(0)}}{g_{t\psi}^{(0)}}\, c =  -  \frac{1}{4p} (3q^2-12 p^2 - 4 F^{(0)} )\, c \,.
\end{equation}
The leading singularity in $g'_{rr}$ in \eqref{rr} can also be eliminated by setting 
\begin{equation}
c = \frac{2 p^2}{\sqrt{3q^2 - 12p^2- 4F^{(0)}}}.
\end{equation}
But there is a further obstruction in eliminating the sub-leading singularity. We want to set 
\begin{equation}
b = \frac{q^3-3p^2 q-F^{(0)} q - F^{(1)} p^2}{p \sqrt{3q^2 - 12p^2- 4F^{(0)}}},
\end{equation}
but again $F^{(1)}$ is generically a function of $\psi$, so we couldn't satisfy this for constant $b$. Thus, we need  to choose the $\rho_I(\psi)$ such that both $F^{(0)}$ and $F^{(1)}$ are constants to have a coordinate transformation which will eliminate the singularities in the metric at $r \to 0$. 
We have three free functions $\rho_1$, $\rho_3$, $\rho_4$, so there should be an arbitrary function's worth of freedom even after satisfying this constraint. We have also seen explicitly above that there are non-trivial solutions with constant $F^{(0)}$ and $F^{(1)}$. 

When we restrict to charge densities such that $F^{(0)}$ and $F^{(1)}$ are constants, the metric near $r=0$ in the new coordinates takes the same general form, 
\begin{equation} \label{newmet2}
ds^2 \approx -\frac{2c}{p} dv dr + f d\hat \psi^2 + 2 h dr d\hat \psi + j(\psi) dr^2 + p^2 (d\theta^2 + \sin^2 \theta d\phi^2), 
\end{equation}
where $f, h$ and $j(\psi)$ are still given by the same general expressions \eqref{fhk}, but now the $g_{rr}$ component $j$ is a function of $\psi$, as $g_{\psi\psi}^{(2)}$ involves $F^{(2)}(\psi)$. In the near-horizon region, $\psi = \hat \psi + c \log(r/p)$, so $j$ is some periodic function which oscillates infinitely many times as we approach the horizon. Thus, while the components of the metric are all finite at the horizon in this new coordinate system, they are not smooth functions of the coordinates there. 

Remarkably, an explicit calculation of the components of the Riemann tensor reveals that all components are finite as $r \to 0$. If we took the approximate metric \eqref{newmet2} to be the exact solution, the only non-zero components of the Riemann tensor are 
\begin{equation}
R_{r \hat \psi r \hat \psi} = -\frac{1}{2} \partial_{\hat \psi}^2 j, \quad R_{\theta\phi \theta\phi} = p^2 \sin^2 \theta,
\end{equation}
which remain finite as $r \to 0$. Keeping sub-leading terms in the expansion in $r$, radial derivatives of $j$ will appear, but they are multiplied by positive powers of $r$, so that we don't get any divergences. Geodesics remain at finite values of the coordinates as $r \to 0$ in \eqref{newmet2}, so the finiteness of the Riemann tensor components implies that the geodesics can be extended beyond $r=0$, so that this is a regular horizon for the oscillating black string solutions. 

Thus, the solution will have a metric which is regular at $r=0$ if we choose the sources such that $F^{(0)}$ and $F^{(1)}$ are constants. There should be a free function's worth of solutions which satisfy these constraints, so we find black string solutions with a function's worth of hair.

\subsubsection{The vector fields}

For the vector fields, we have
\begin{equation} \label{vector}
A^I  = Z^{-1} X^I  \left ( dv - \frac{a}{r^2} dr - \frac{b}{r} dr \right) + \left( Z^{-1} X^I  \mu  - K^I \right) \left( d \hat \psi + \frac{c}{r} dr \right)  + p \cos \theta d\phi + \alpha^I \,,
\end{equation}
with $ d\alpha^I =0$.
Again, these components are all finite in the original coordinate system, but we need to check carefully that they are finite in the regular coordinate system as $r \to 0$. For $I=0$, 
\begin{equation}
Z^{-1} X^0  \mu - K^0 \approx \frac{q}{2p}, 
\end{equation}
so we take $\alpha^0 = - \frac{q}{2p} d\psi$ as before. For $I=1,2$,
\begin{equation}
Z^{-1} X^I  \mu - K^I \approx \frac{q}{2p} \mp \frac{\rho_1(\psi)}{p}.
\end{equation}
Taking $\alpha^I = - \frac{q}{2p} d\psi$ will eliminate the first term as before.  Since we take the charge distributions to be purely oscillating, the new term is actually pure gauge, so it can be removed by taking  $\alpha^I =- \frac{q}{2p} d\psi+  d \beta^I(\psi)$ for some suitably chosen periodic functions, $\beta^I(\psi)$.  Similarly for $I =3,4$, 
\begin{equation}
Z^{-1} X^I  \mu  \approx - \frac{\rho_I(\psi)}{p}, 
\end{equation}
which can  also be eliminated by a gauge transformation. All the vector potentials are then finite as $r \to 0$. 

There is a subtlety here similar to the one we saw above in the metric. The $A_v$ and $A_{\hat \psi}$ components vanish at the horizon after we do this gauge transformation, so the surviving components at the horizon are $A_\phi = p \cos \theta$ and $A_r$. The finite contribution to $A_r$ from the first term in \eqref{vector} is independent of $\psi$, but the contribution from the second term is
\begin{equation}
A_r \sim c \left( \frac{\partial_r F_I(\psi, r)|_{r=0}}{p}  - \frac{q \rho_I(\psi)}{2p^3} \right).
\end{equation}
This is bounded, but as $\psi = \hat \psi + c \log (r/p)$, it will  oscillate infinitely many times along infalling geodesics which approach the horizon at finite $\hat \psi$. Since it is only the radial component of the gauge field that exhibits this behaviour, it does not lead to a divergent field strength; the non-zero components of the field strength at the horizon are  
\begin{equation}
F_{\hat \psi r} = \partial_{\hat \psi} A_r, \quad F_{\theta \phi} = p \sin \theta.
\end{equation}
Thus, the electric field is bounded but  oscillates infinitely many times along infalling geodesics approaching $r=0$. Thus, the response of a charged test particle remains bounded, and the horizon at $r=0$ remains regular\footnote{We can see that derivatives of the field strength will diverge at the horizon, so the field is not smooth there, but it is sufficiently differentiable to admit a physically meaningful extension through the horizon.}.

\section{Oscillating black rings} 
\label{Sect:Orings}

We now extend the discussion to the most interesting example, an oscillating black ring. As we remarked earlier, since the smooth continuation of a solution across the horizon is a local issue we expect to encounter only the same issues and constraints as we did for the black string, albeit with a slightly higher level of complexity.

\subsection{The O-ring solution} 

To bend our black string solution into a black ring, we want to take the four-dimensional base space \eqref{GHmetric} to be flat $\mathbb R^4$. This is achieved by introducing a single GH centre at $r_1 =0$ and setting $V = 1/r_1$. Note that it is rather straightforward to extend our construction to oscillating rings in Taub-NUT or more complicated multi-center solutions, although we will not do it here for the sake of simplicity. We will choose coordinates so that the black ring's event horizon is still at $r=0$, and take the centre $r_1 = 0$ to be at  $r=R$, $\theta=0$. That means 
\begin{equation}
r_1^2 = r^2 + R^2 - 2rR \cos \theta .
\end{equation}
We have  
\begin{equation}
V= \frac{1}{r_1} = \frac{1}{\sqrt{r^2 + R^2 - 2rR \cos \theta}}, \quad A = \frac{ r_1 + r \cos \theta - R}{r_1} d\phi. 
\end{equation}

We take the harmonic functions $K^I$, $L_I$ to have the same structure as in the black string solution, given in \eqref{kfns}, \eqref{lfns}. The $L_I$ satisfy $\nabla_4^2 L_{I} = (q+ \rho_I(\psi)) \delta^3(\vec x)$, which now implies 
\begin{equation} \label{GF}
F_I(r, \theta, \psi) = r \int_0^{4\pi} G(r, \theta, \psi, \psi') \rho_I(\psi') d\psi'
\end{equation}
where $G$ is the Page Green function for the GH base \cite{Page:1979ga,Bena:2010gg}. For the single-centred base, this Green function is simply 
\begin{equation}
G  = \frac{1}{16 \pi^2 r} \frac{\sinh U}{\cosh U - \cos(\frac{\psi - \psi'}{2}) },
\end{equation}
where $2U = \ln \frac{r_1 + R + r}{r_1 + R - r}$. Note that the $\theta$ dependence in $r_1$ implies 
that $F_I$ is a function of $\theta$. As for the black string, we can do the integral in \eqref{GF} for appropriate sources by contour integration, to get 
\begin{equation} \label{anF}
F_I(r, \theta, \psi) = \mathrm{Re} [ \rho_I(\psi + i U)] \approx \rho_I (\psi) + \frac{r}{2R} \mathrm{Re} [i \partial_z \rho_I (z)|_{z=\psi}] + \mathcal{O} (r^2) 
\end{equation}
The functions $Z_I$ are now given by:
\begin{equation}
Z_0 = 1 + \frac{q}{r} + \frac{p^2}{r^2 V}, \quad Z_{1,2} = 1 + \frac{q}{r} \pm \frac{F_1}{r} + \frac{p^2}{r^2 V}, \quad Z_{3,4} = \frac{F_{3,4}}{r}, 
\end{equation}
and so the warp factor appearing in the metric is:
\begin{equation}
Z^3 = Z_0 (Z_1 Z_2 - \coeff{1}{2}( Z_3^2+  Z_4^2)) = \left( 1 + \frac{q}{r} + \frac{p^2}{r^2 V} \right) \left(  \left( 1 + \frac{q}{r} + \frac{p^2}{r^2 V} \right)^2 - \frac{(F_1^2 +\frac{1}{2}( F_3^2+ F_4^2))}{r^2} \right).
\end{equation}
We also have
\begin{equation}
\mu = \frac{p^3}{r^3 V^2} + \frac{3pq}{2r^2 V} + \frac{3p}{2 r V} + M. 
\end{equation}
As before, this part of the solution does not involve the oscillation. We therefore choose $M$ to have the same value as for the non-oscillating black ring,
\begin{equation} 
M = \frac{3p}{2} \frac{R-r}{r}
\end{equation}
The one-form is then also the same as for the non-oscillating ring, 
\begin{equation}
\omega_\phi  = - \frac{3p  R}{2} \frac{(r -R \cos \theta + r_1 \cos \theta)}{R\, r_1} + \frac{3p}{2} A_\phi.
\end{equation}
Again, dependence on the oscillation enters in the metric only through $Z$. 

As before, we can find a simple solution in which the metric is completely independent of $\psi$, by choosing 
\begin{equation}
\rho_1(\psi) = - \rho_2(\psi) =  Q \cos (k \psi), \quad \rho_3  = \rho_4 = Q \sin (k \psi), 
\end{equation}
so that \eqref{anF} gives 
\begin{equation}
\label{simpexamp1}
F_1 = Q  \cos k \psi \, e^{-k U}, \quad F_3 = F_4 = Q  \sin k \psi \, e^{-k U} ~\Rightarrow~ F_1^2 + \coeff{1}{2}(F_3^2 + F_4^2) = Q^2 e^{-2kU}, 
\end{equation}
Thus $\frac{\partial}{ \partial \psi}$  is a Killing vector. Note that the warp factor is still a function of $\theta$, unlike the black-string solution. The form of $U$ implies that the oscillation now has power-law decay at large $r$, as expected for multipole moments in an asymptotically-flat space. 

More generally, if the source functions were arbitrary, the $F^I$ would have power-law decay at large $r$, and one could use eq. \eqref{anF} to see that the expansion around $r=0$ becomes
\begin{equation}
\label{Fdefns}
F_1^2 + \frac{1}{2} (F_3^2 + F_4^2) = F^{(0)} (\psi) + F^{(1)}(\psi)\, r + F^{(2)}(\theta,\psi)\,  r^2 + \ldots
\end{equation}
A key point is that the $\theta$ dependence enters only at quadratic order; near $r=0$, $U \approx r/2R$, so the linear term in the expansion of $F_I$ is still $\theta$-independent, as indicated in \eqref{anF}. 

\subsection{Regularity}

The regularity analysis near $r=0$ now proceeds in much the same manner  as it did  for the black string.

\subsubsection{The metric and scalars}

 The scalars take the same form \eqref{scalars}, and as we approach $r=0$, $X^I \to 1$ for $I=0,1,2$ and $X^I \to 0$ for $I=3,4$. For the metric, we note that as $r \to 0$,
\begin{equation}
V \approx \frac{1}{R}, \quad A \approx \frac{r^2}{2R^2} \sin^2 \theta d\phi, \quad \omega \approx - \frac{3p}{2} r \sin^2 \theta d\phi, 
\end{equation}
so the cross terms involving $\phi$ become negligible near $r=0$. The metric near $r=0$ is then
\begin{equation}
ds^2  \approx -\frac{r^4}{p^4 R^2} dt^2  -\frac{2r}{p} dt d \psi + \frac{1}{4 p^2} (3 q^2 - 12 p^2 R - 4 F^{(0)})  d\psi^2  + \frac{p^2}{r^2} dr^2 + p^2 (d\theta^2 + \sin^2 \theta d\phi^2).   
\end{equation} 
If we choose charge densities such that  the sum of the squares $F^{(0)}(\psi) = \rho_1^2 + \frac{1}{2} (\rho_3^2 + \rho_4^2)$ is a constant, we can eliminate the potential singularity in $g'_{r\psi}$ in \eqref{rpsi} by setting 
\begin{equation}
a =  \frac{g_{\psi\psi}^{(0)}}{g_{t\psi}^{(0)}} \, c =  - \frac{c}{4 p} (3 q^2 - 12 p^2 R - 4 F^{(0)}), 
\end{equation}
and the leading singularity in $g'_{rr}$ in \eqref{rr} can also be eliminated choosing $c$ appropriately. We also need to choose sources such that $F^{(1)}(\psi)$ is also a constant, so that we can eliminate the sub-leading singularity in $g'_{rr}$ by choosing $b$ appropriately. 

When we restrict to charge densities such that $F^{(0)}$ and $F^{(1)}$ are constants, the metric near $r=0$ in the new coordinates takes the same general form, 
\begin{equation} \label{newmet3}
ds^2 \approx -\frac{2c}{p} dv dr + f d\hat \psi^2 + 2 h dr d\hat \psi + j(\theta,\psi) dr^2 + p^2 (d\theta^2 + \sin^2 \theta d\phi^2), 
\end{equation}
where again $f, h$ and $j(\theta,\psi)$ are given as in \eqref{fhk}, but now  $j$ is a function of both $\theta$ and $\psi$, as $g_{\psi\psi}^{(2)}$  contains $F^{(2)}(\theta,\psi)$. As before, in the near-horizon region, $\psi = \hat \psi + c \log(r/p)$, so $j$ is some periodic function that oscillates infinitely many times as we approach the horizon. Thus, while the components of the metric are all finite at the horizon in this new coordinate system, they are not smooth functions of the coordinates there. 

Remarkably, it is still true that the components of the Riemann tensor in the approximate metric \eqref{newmet2} are finite as $r \to 0$ even after including the $\theta$ dependence. The leading metric now has the Riemann tensor
\begin{equation}
R_{r\hat{\psi} r \hat \psi} = - \frac{1}{2} \partial_{\hat \psi}^2 j, \quad R_{r \hat \psi r \theta} = - \frac{1}{2} \partial_\theta \partial_{\hat \psi} j, \quad R_{r \theta r \theta} = - \frac{1}{2} \partial_\theta^2 j, \quad R_{r \phi r \phi} = - \frac{1}{2} \partial_\theta j \sin \theta \cos \theta, 
\end{equation}
\begin{equation}
R_{\theta \phi \theta \phi} = p^2 \sin^2 \theta, \nonumber
\end{equation}
which is finite as $r \to 0$.  Including sub-leading terms in the metric, radial derivatives of $j$ again appear multiplied by positive powers of $r$, so  we don't get any divergences in the Riemann tensor. Geodesics remain at finite values of the coordinates as $r \to 0$ in \eqref{newmet2}, so the finiteness of the Riemann tensor components implies that the geodesics can be extended beyond $r=0$, so that this is a regular horizon for the oscillating black ring solutions.  

Thus, the solution will have a metric that is regular at $r=0$ if we choose the sources such that $F^{(0)}$ and $F^{(1)}$ are constants. Since these are in general functions only of $\psi$, this imposes two restrictions on our three free source functions, so there should be a free function's worth of solutions that satisfy these constraints.  Thus we find black ring solutions with a function's worth of hair. It is essential that the regularity only requires constancy of $F^{(0)}$ and $F^{(1)}$; requiring constancy of $F^{(2)}$, or even higher order terms, which are  functions of both $\theta$ and $\psi$ in general, would  overconstrain the sources. 

\subsubsection{The vector fields}

The analysis for the vector fields is essentially identical to that for black strings. The fields are: 
\begin{equation}
A^I  = Z^{-1} X^I  \left ( dv - \frac{a}{r^2} dr - \frac{b}{r} dr \right) + \left( Z^{-1} X^I  \mu  - \frac{K^I}{V} \right) \left( d \hat \psi + \frac{c}{r} dr \right)  + p \cos \theta d\phi + \alpha^I.
\end{equation}
The components in the new coordinates are made finite by a suitable choice of exact part. For $I=0$,  as $r \to 0$
\begin{equation}
Z^{-1} X^0  \mu - \frac{K^0}{V} \approx \frac{q}{2p}, 
\end{equation}
so we take $\alpha^0 = - \frac{q}{2p} d\psi$ as before. For $I=1,2$,
\begin{equation}
Z^{-1} X^I  \mu - \frac{K^I}{V} \approx \frac{q}{2p} \mp \frac{\rho_1(\psi)}{p}\,,
\end{equation}
and the divergence of the vector field can be cancelled by a gauge transformation and taking  $\alpha^I = - \frac{q}{2p} d\psi + d \beta^I$ for a suitable choice of $\beta^I(\psi)$.  Similarly for $I =3,4$, 
\begin{equation}
Z^{-1} X^I  \mu  \approx - \frac{\rho_I(\psi)}{p}, 
\end{equation}
which can be eliminated by a gauge transformation involving the choice of $\beta^I(\psi)$. All the vector components are then finite as $r \to 0$. As for black strings, there is a finite $A_r$ component that is a function of $\psi = \hat \psi + c \log (r/p)$,\footnote{This comes from the linear term in $F_I$, so it is a function just of $\psi$, and not of $\theta$.} which gives an electric field that is bounded but oscillates infinitely many times along infalling geodesics. Thus, the response of a charged test particle remains bounded, and the horizon at $r=0$ remains regular.


\section{The solution in canonical ring coordinates}
\label{CanonCoords}

We can easily recast the solutions above in the, perhaps more familiar, black-ring bipolar coordinates. This will facilitate the comparison with the singular black ring solution that only has two oscillating charge densities \cite{Bena:2004td}, and to this end we also modestly generalize the result above by using distinct magnetic dipole moments.
  
\subsection{The oscillating solutions}

One can write the $\IR^4$  spatial base metric of (\ref{metAnsatz}) in the usual spherical bipolar coordinates:
\begin{equation} 
ds^2_{\IR^4} ~=~ {R^2 \over (x-y)^2}\, \bigg( {dy^2 \over y^2 -1} 
+ (y^2-1)\, d \psi^2  +{dx^2 \over 1-x^2} + (1-x^2) \, d \phi^2  \bigg)\,.
\end{equation}
with  $ -1 \le x \le 1$, $-\infty < y \le -1$.  The  ring is located at $y = -\infty$ while spatial infinity corresponds to $y=-1$.

In these coordinates the magnetic flux sources are very simple:
\begin{equation} 
\Theta^{(j)} ~=~  2\, q_j \, (d x \wedge d \phi ~-~ d y \wedge d \psi) \,,\qquad j=0,1,2  \,
\end{equation}
and the electrostatic potentials that solve \ref{BPSeqn:2} have the form \cite{Bena:2004td}:
\begin{align} 
Z_i  ~=~ & 1 ~+~{Q_i \over R }\, (x-y)  ~-~ {4 q_j \, q_k \over R^2} \, 
(x^2 - y^2)   ~+~  {\pi \over R }\, (x-y)\,  \Lambda_i(y,\psi) \,, \quad \{i,j,k\} = \{0,1,2\} \,,\\
Z_i  ~=~ &  {\sqrt{2} \, \pi \over R }\, (x-y)\,  \Lambda_i(y,\psi) \,, \quad  i = 3,4  \,, 
\end{align}
where, for later convenience, we have introduced factors of $\sqrt{2}$ into $\Lambda_3$ and $\Lambda_4$.  The  $\Lambda_i$ are harmonic, which means that they satisfy:
\begin{equation} 
(y^2 -1) \,  \partial_y \big( (y^2 -1) \,  \partial_y \, \Lambda_i \big)
~+~ \partial_\psi^2 \,  \Lambda_i ~=~ 0 \,,
\end{equation}
The Fourier analysis is elementary and the smooth solutions that fall off at infinity have expansion: 
\begin{equation} 
\Lambda_i(y,\psi) ~\equiv~  
\sum_{n=1}^ \infty \, \Big({y+1 \over y-1} \Big)^{n/2} \,  \big(\, a_n^i \, \cos(n\,\psi) ~+~ b_n^i \, \sin(n\,\psi)\,\big) \,.
\end{equation}
Indeed, if one takes $y = -\coth \xi$ then this may be written as the real part of an analytic function  of $z$ where $z \equiv e^{-\xi + i \psi}$.

Note that we focus on solutions where the additional potentials, $Z_3$ and $Z_4$, only have a fluctuating piece. Moreover, the source in the third BPS equation (\ref{BPSeqn:3}) only involves the sum 
\begin{equation} 
\Lambda ~\equiv ~ \sum_{i=0}^ 2 \, q_i\, \Lambda_i\,.
\end{equation}
The solution to the third BPS equation (\ref{BPSeqn:3})  with generic fluctuations  was given in \cite{Bena:2004td} however, our purpose here is to minimize, or at least suppress, the $\psi$-dependence of the metric. We will therefore consider charge densities that fluctuate while keeping $\Lambda \equiv 0$.  We also want to try to remove the $\psi$-dependence from $Z$, which is given by (\ref{Zreln2}) and therefore consider a solution where the charge corresponding to $\Lambda_0$ does not fluctuate and thus $\Lambda_0 \equiv 0$ as well\footnote{There may be more general solutions but our purpose here is to exhibit some simple families and leave the classification issue for later work.}.  Thus we have 
\begin{equation} 
\Lambda_0 ~\equiv~ 0 \,, \qquad q_1 \Lambda_1 ~+~  q_2 \Lambda_2~\equiv~ 0 \,.
\label{chgconstr1}
\end{equation}
With these choices, the angular momentum vector, $k$, is precisely what it was for the non-oscillating black ring:
\begin{equation} 
k ~=~ k_1 \, d\psi ~+~ k_2 \, d \phi \,,
\end{equation}
where 
\begin{align} 
 k_1 ~=~  & (y^2-1)\,\big(\, \coeff{1}{3}\,C \,(x+y) ~+~
\coeff{1}{2} \,B \big) ~-~ A\, (y+1) \,,  \\   
k_2 ~=~ &  (x^2-1)\,\big(\, \coeff{1}{3}\,C \,(x+y) ~+~
\coeff{1}{2} \,B \big)    \,,
\end{align} 
and
\begin{equation} 
A ~\equiv~  2 (q_0+q_1+q_2)\,, \quad B ~\equiv~ 
{2\over R} (Q_0\, q_0 + Q_1\, q_1 + 
Q_2\, q_2)\,, \quad  C~\equiv~ -  {24 q_0\, q_1\, q_2 \over R^2}\,.
\end{equation}

Thus the $\psi$-dependence has almost been eliminated from the five-dimensional metric. Indeed, the only such dependence comes from the warp factors (\ref{Zreln2}) via (\ref{Pdefn}):
\begin{equation}
P   ~=~ \big( Z_1 \, Z_2  - \coeff{1}{2} \, (Z_3^2  + Z_4^2) \big)\,.   
\end{equation}
The fluctuating part, $\widehat P$,  of $P$ is given by:
\begin{align}
\widehat P   ~=~ & \frac{\pi}{R}\, (x-y)\, \big(\Lambda_1  ~+~ \Lambda_2 \big) ~+~  \frac{\pi}{R^2}\, (x-y)^2\, \big(Q_2 \, \Lambda_1 + Q_1 \, \Lambda_2\big) \\
 & ~+~  \frac{\pi^2}{R^2}\, (x-y)^2\, \big( \Lambda_1  \, \Lambda_2  -  (\Lambda_3^2  + \Lambda_4^2) \big)\,,  
\end{align}
where we have used (\ref{chgconstr1}).

\subsection{An elementary example}
\label{Sect:Elem}

Take $q_1 = q_2$  and $Q_1 = Q_2$ then $\Lambda_2 = -\Lambda_1$ and 
\begin{equation}
\widehat P   ~=~   - \frac{\pi^2}{R^2}\, (x-y)^2\, \big( \Lambda_1^2 + \Lambda_3^2  + \Lambda_4^2 \big)\,,  
\end{equation}

We can take the fluctuations to be a single Fourier mode:
\begin{equation} 
\Lambda_i(y,\psi) ~=~   \Big({y+1 \over y-1} \Big)^{n/2} \,  \big(\, a_n^i \, \cos(n\,\psi) ~+~ b_n^i \, \sin(n\,\psi)\,\big) \,,
\label{Fourier}
\end{equation}
and then one has
\begin{equation} 
\widehat P   \! ~=~ \!    - \frac{\pi^2}{2\, R^2}\, (x-y)^2\, \Big({y+1 \over y-1} \Big)^{n} \!\!\! \sum_{i\in \{1,3,4\}} \!\!\! \big[\, \big((a_n^i)^2 + (b_n^i)^2\big) ~+~ \big((a_n^i)^2 - (b_n^i)^2\big) \, \cos(2 n\psi) ~+~   2\, a_n^i \, b_n^i\,\sin(2n \psi) \,\big].
\end{equation}
Consider the $a_n^i $ and $b_n^i $ to be vectors, $\vec a, \vec b \in \IR^3$, then the fluctuations in the metric are completely absent if
\begin{equation} 
|\vec a |^2  ~=~ |\vec b |^2  \,, \qquad   \vec a \cdot \vec b ~=~0   \,.
\label{abconstr}
\end{equation}
That is, the coefficients are orthogonal vectors of the same length in $\IR^3$ and then:
\begin{equation} 
\widehat P   ~=~     - \frac{\pi^2\, |\vec a |^2 }{R^2}\, (x-y)^2\, \Big({y+1 \over y-1} \Big)^{n}  \,.
\end{equation}

This is simply a modest extension of the example given in (\ref{simpexamp1}).   Here we have a four-parameter family of solutions:  $\vec a$ can be freely chosen and then $\vec b$ is vector of the same length in the plane orthogonal to $\vec a$. Thus there is really {\it a three parameter family of hair} because only $|\vec a |$ appears in the metric. 

There is a simple way to parametrize the solutions of (\ref{abconstr}) by setting
\begin{equation} 
\vec a + i \vec b  ~=~  (i(\zeta_1^2 + \zeta_2^2) \,, \,\zeta_1^2 - \zeta_2^2\,, \, 2\,\zeta_1\, \zeta_2)  \,.
\label{quadric}
\end{equation}
for any $(\zeta_1,\zeta_2) \in \IC^2$.  This automatically satisfies  (\ref{abconstr}) and $|\vec a |^2 + |\vec b |^2  = 2(|\zeta_1|^2 + |\zeta_2|^2)$.  We therefore see that the solution space for fixed $\widehat P$ is $S^3/\ZZ_2 = SO(3)$ where the $\ZZ_2$ action is simply $(\zeta_1,\zeta_2)  \to -(\zeta_1,\zeta_2)$.

In this example, $\frac{\partial}{\partial \psi}$ is a Killing vector of the metric but not a symmetry of the solution.  These solutions resemble therefore $Q$-balls, in that the oscillations cancel out of the energy-momentum tensor and the metric, but the charges rotate around the ring. 

\subsection{Horizon regularity in general}

We can check horizon regularity  exactly as we did earlier. One first makes a change of variable $u= - 1/y$ and then tries to  continue  smoothly through $u=0$.  We will need the expansion, to first order,  of the electric field terms, $\Lambda_i$, in the neighborhood of the horizon  ($y=-\infty$): 
\begin{equation} 
\Lambda_i (y, \psi ) ~=~\Lambda^{(0)}_i (\psi)  + \Lambda^{(1)}_i (\psi) \, y^{-1} ~+~ \dots \,.
\end{equation}
To remove the singularities one must also allow $u$-dependent shifts in the angular and time coordinates:
\begin{equation} 
t ~=~  v +\frac{a}{u} + b \, \log u \,, \qquad \psi ~=~  \varphi_1  + c_1 \, \log u \,, \qquad   \phi ~=~  \varphi_2  + c_2 \, \log u \,,
\label{varchng}
\end{equation}
where $a$, $b$,$c_1$ and $c_2$ are constants to be determined.   

It is convenient to define two quantities:
\begin{align} 
\Delta_1 ~\equiv~& 4 \pi^2 \, q_1\, \big( q_1  (\Lambda^{(0)}_1)^2 + q_2  ((\Lambda^{(0)}_3)^2+(\Lambda^{(0)}_4)^2)\big)  +4 \pi  \, q_1  (q_1 Q_1  -q_2 Q_2) \Lambda^{(0)}_1   \,, \\
\Delta_2 ~\equiv~& 4 \pi^2  \, q_1\,  \Big( q_1  \,\Lambda^{(0)}_1 \, \Lambda^{(1)}_1  + q_2  \, \Big(  \Lambda^{(0)}_3 \, \Lambda^{(1)}_3 +    \Lambda^{(0)}_4 \, \Lambda^{(1)}_4\Big)  \Big)  \\
& \qquad \qquad\qquad\qquad + 2 \pi  \, q_1\,  (q_1 Q_1  -q_2 Q_2) \,  \Lambda^{(1)}_1  -  2 \pi \, R\, q_1\, (q_1 - q_2)\,\Lambda^{(0)}_1   \,.
\end{align}
These quantities are in principle $\psi$-dependent, but as we will see, smoothness across the horizon requires that these quantities are in fact independent of $\psi$.

One then expands  the metric about $u=0$.  
To remove the $u^{-1}$ terms  proportional to $du  \,d\varphi_2$ one must set $c_2 = - c_1$.  The  $u^{-1}$ terms  proportional to $du  \, d\varphi_1$  can be removed by setting 
\begin{equation} 
a ~=~  \frac{8\, q_0 q_1 q_2}{R^2} \,  \Omega^2 \, c_1\,,
\label{aval}
\end{equation}
where
\begin{align} 
 \Omega ~\equiv~ \frac{R}{8\, q_0 q_1 q_2} \, \Big[  & 2 \,(q_0 q_1 Q_0 Q_1+ q_0 q_2 Q_0 Q_2+q_1 q_2 Q_1 Q_2) \\
&- 16\, q_0 q_1 q_2 (q_0 + q_1 + q_2) - (q_0^2 Q_0^2 + q_1^2 Q_1^2 + q_2^2 Q_2^2)  - \Delta_1\Big]^{1/2} \,.
\label{Omdefn}
\end{align}
To remove the $u^{-2} du^2$ terms  one must set 
\begin{equation} 
c_1 ~=~  \frac{1}{\Omega} \,.
\label{c1val}
\end{equation}
 The $u^{-1} du^2$ terms have pieces that are linear in $x$ but these cancel as a consequence of (\ref{aval}) and (\ref{c1val}).  This is the equivalent in the $(x,y)$ coordinates of the $\theta$-independence of $F^{(0)}$ and $F^{(1)}$ in (\ref{Fdefns}).  The remaining terms involving $u^{-1} du^2$ can be removed by setting:
\begin{align} 
b ~=~ -\frac{1}{\Omega} \, \bigg[  & \frac{1}{16\, q_0 q_1 q_2} \, \Big(  32 q_0 q_1 q_2 (q_0 + q_1 + q_2)   - 4 R\,  (q_0^2 Q_0 + q_1^2 Q_1 + q_2^2 Q_2) +R\, Q_0 Q_1 Q_2  \Big)  \\
& ~-~ \frac{R\, q_0 Q_0\, \Delta_1}{16\, q_0^2 q_1^2 q_2^2}  ~+~  \frac{1}{8\, q_0 q_1   q_2} \,  \Delta_2\bigg]  \,.
\label{bval}
\end{align}
With this change of variables, the metric continues smoothly across $u=0$.  (If the $\Delta_i$ depended on $\psi$ then the change of variables (\ref{varchng}) would have introduced new singular $d \psi$ terms coming the derivatives of $\Delta_i$.)  We have not verified that the other fields and the Riemann tensor are finite across the horizon in these coordinates, but this is guaranteed by the analysis in Section \ref{Sect:Orings}.

Note that if we choose $q_1 = q_2$ and $Q_1 = Q_2$ then we should recover the same solution as in Section \ref{Sect:Orings}.  Indeed, requiring that $\Delta_1$ and $\Delta_2$ be constant reduces to requiring the $\psi$-independence of 
\begin{equation} 
\sum_ {i \in \{1,3,4\}}  (\Lambda^{(0)}_i (\psi))^2     \qquad {\rm and} \qquad \sum_ {i \in \{1,3,4\}}   \Lambda^{(0)}_i (\psi) \, \Lambda^{(1)}_i (\psi)   \,.
\end{equation}
This is equivalent to requiring that the following are $\psi$-independent:
\begin{equation} 
\sum_ {i \in \{1,3,4\}}  (\Lambda_i (y,\psi))^2  \Big |_{y=-\infty}   \qquad {\rm and} \qquad \lim_{y\to -\infty} \,\Big [ y^2 \frac{\partial}{\partial y}\,  \sum_ {i \in \{1,3,4\}}  (\Lambda_i (y,\psi))^2\Big]  \,.
\end{equation}
This is precisely the same condition as requiring $F^{(0)}$ and $F^{(1)}$ in (\ref{Fdefns}) be independent of $\psi$.

Given these two functional constraints, we can ask how big will be the family of hair that our solutions can accommodate. We started with four charge density functions constrained by (\ref{chgconstr1}), and thus are three unconstrained functions parameterized by the arbitrary Fourier coefficients $a_n^i$ and $b_n^i$, $i \in \{1,3,4\}$ in (\ref{Fourier}). Imposing the two extra functional  constraints we found should then leave, in principle, one function's-worth of hair on the surface of the black ring. Although we did not find a simple way to prove that this will always happen, it seems very plausible particularly because we can find a simple explicit example (see Section \ref{Sect:Elem}) and thus show that the set solutions is certainly  non-empty.  We therefore believe our analysis provides good evidence that there is a lot of new oscillatory hair intrinsic to black O-rings and thus to black rings in M-theory.

\section{Conclusions} 

We have found some very simple BPS black-ring solutions that exhibit whole new varieties of hair through charge oscillations.  The existence of these O-ring solutions requires that we go beyond the usual three-charge, or ``STU,'' supergravity and couple more vector multiplets.  It is only in this way that we can sufficiently expand the phase space of solutions to have enough freedom to coiffure the hair so as to arrange smoothness across the horizon.

We have not attempted to provide a complete classification of these new BPS O-ring solutions. There are obviously more general possibilities, even within the supergravity models that we have considered here.  The first step  was to make sure that all the fluctuations cancel in the source term of (\ref{BPSeqn:3}) so that the angular momentum vector remained $\psi$-independent and thus the only $\psi$-dependence in the metric then comes through the warp factor, $Z$.    With more non-zero independent magnetic dipoles there are manifestly many ways to arrange this in the source of (\ref{BPSeqn:3}).    For simplicity, we also suppressed fluctuations in $Z_0$. Moreover, generic Calabi-Yau manifolds give rise to many more vector fields and more complicated intersection matrices. There is therefore a very large  phase space that could be explored by O-rings in string theory. 

In terms of counting solutions we have argued that, even within the limited class considered here, there appears to be a whole function's worth of hair, essentially because we have three charge density functions and two functional constraints coming from requiring smoothness at the horizon.  While this is not a formal proof and  some subtlety might yet appear in a more careful analysis, our construction strongly suggests that such large families of  very hairy solutions exist. They would give by far the largest known violation of black hole uniqueness: an arbitrary function of one variable is parameterized by an infinite number of arbitrary Fourier modes, and hence our solution would depend on an infinite number of continuous parameters.

Even if somehow these functional constrains are too strong and do not give a coiffure parameterized by an infinite number of continuous parameters, we have explicitly exhibited a three-parameter family of hair on BPS O-rings. This in itself provides a substantial enlargement of the hairiness of black objects. 

One rather surprising aspect of our work is that the curvatures and gauge fields are continuous across the horizon.  The functional constraints on the charge densities manifestly arrange that the metric is continuous and non-degenerate across the horizon.  However in the  coordinates that continue across the horizon, the warp factor, $Z$, depends upon oscillations whose frequency diverges as one approaches the horizon.  Of course, as a function of the original coordinates, the behavior of $Z$ is completely well-behaved and it is only through transforming to the infalling coordinates that one sees such infinite frequencies.  This is presumably because such an infalling observer rotates around the ring infinitely often before crossing the horizon and so sees the charge oscillations with diverging frequency.  The surprise is that in spite of this behavior, the curvature and field strengths are not divergent in crossing the horizon and test particles do not encounter infinite fields.  The higher level of smoothness is evidently related to the fact that the oscillations of the metric are limited entirely to the five-dimensional warp factor and that its leading two orders of oscillation have been eliminated by the charge density constraints.  However, the finiteness of the curvatures and gauge fields is still surprising and we hope to investigate this further.

It seems very plausible that the ideas employed here could be extended to extremal non-BPS\footnote{The near-horizon geometry of extremal non-BPS black rings \cite{Bena:2009ev} appears to be more complicated than that of BPS rings \cite{Bossard:2012ge}, but since the underlying equations are still linear it may be possible to ensure smoothness by suitable choosing the oscillating charge distributions.} and even non-extremal black rings \cite{Elvang:2004xi}.  Indeed, one of the inspirations for this work was the idea that one might hope to find non-BPS microstate geometries that oscillate in the manner described here and would ultimately be electrically neutral \cite{Mathur:2013nja}.   One of the other inspirations for this work were  the Q-ball solutions  \cite{Coleman:1985ki,Friedberg:1986tq} which are {\it time}-dependent and thus certainly not BPS.  One might hope to find time-dependent O-rings and even time-dependent microstate geometries that oscillate in the same manner.  The problem with such time-dependent solutions within string theory is that such charge densities will generically couple to the Maxwell fields are thus emit electromagnetic radiation.  It is, or course, possible that one might find a way to do this sufficiently coherently so as to suppress the radiation within a suitably complicated supergravity phase space.  


\bigskip
\leftline{\bf Acknowledgements}
\smallskip
NPW is grateful to the IPhT, CEA-Saclay and to the Institut des Hautes Etudes Scientifiques (IHES), Bures-sur-Yvette, for hospitality while this work was initiated. NPW would also like to thank the Simons Foundation for their support through a Simons Fellowship in Theoretical Physics. We are all grateful to the Centro de Ciencias de Benasque for hospitality at the ``Gravity- New perspectives from strings and higher dimensions'' workshop. The work of IB was supported in part by the ERC Starting Independent Researcher Grant 240210-String-QCD-BH, by the John Templeton Foundation Grant 48222: ``String Theory and the Anthropic Universe'' and by a grant from the Foundational Questions Institute (FQXi) Fund, a donor advised fund of the Silicon Valley Community Foundation on the basis of proposal FQXi-RFP3-1321 to the Foundational Questions Institute. The work of NPW  was supported in part by the DOE grant DE-FG03-84ER-40168. The work of SFR was supported in part by the STFC. 



\begin{thebibliography}{99}

\bibitem{Mateos:2001qs} 
  D.~Mateos and P.~K.~Townsend,
  ``Supertubes,''
  Phys.\ Rev.\ Lett.\  {\bf 87}, 011602 (2001)
  [hep-th/0103030].

\bibitem{Lunin:2001fv} 
  O.~Lunin and S.~D.~Mathur,
  ``Metric of the multiply wound rotating string,''
  Nucl.\ Phys.\ B {\bf 610}, 49 (2001)
  [hep-th/0105136].

\bibitem{Lunin:2002iz} 
  O.~Lunin, J.~M.~Maldacena and L.~Maoz,
  ``Gravity solutions for the D1-D5 system with angular momentum,''
  hep-th/0212210.

\bibitem{Bena:2004de}
  I.~Bena and N.~P.~Warner,
  ``One ring to rule them all ... and in the darkness bind them?,''
  Adv.\ Theor.\ Math.\ Phys.\  {\bf 9} (2005) 667
  [hep-th/0408106].
    
\bibitem{Elvang:2004ds} 
  H.~Elvang, R.~Emparan, D.~Mateos and H.~S.~Reall,
  ``Supersymmetric black rings and three-charge supertubes,''
  Phys.\ Rev.\ D {\bf 71}, 024033 (2005)
  [hep-th/0408120].
    
\bibitem{Gauntlett:2004qy} 
  J.~P.~Gauntlett and J.~B.~Gutowski,
  ``General concentric black rings,''
  Phys.\ Rev.\ D {\bf 71}, 045002 (2005)
  [hep-th/0408122].
  
      
\bibitem{Bena:2004td} 
  I.~Bena, C.~-W.~Wang and N.~P.~Warner,
 ``Black rings with varying charge density,''
  JHEP {\bf 0603}, 015 (2006)
  [hep-th/0411072].
  
  \bibitem{Horowitz:2004je}
  G.~T.~Horowitz and H.~S.~Reall,
  ``How hairy can a black ring be?,''
  Class.\ Quant.\ Grav.\  {\bf 22} (2005) 1289
  [hep-th/0411268].
  
  \bibitem{Bena:2013ora}
  I.~Bena, S.~F.~Ross and N.~P.~Warner,
  ``On the Oscillation of Species,''
  arXiv:1312.3635 [hep-th].
  
\bibitem{Coleman:1985ki} 
  S.~R.~Coleman,
``Q Balls,''
  Nucl.\ Phys.\ B {\bf 262}, 263 (1985)
  [Erratum-ibid.\ B {\bf 269}, 744 (1986)].
 
\bibitem{Friedberg:1986tq} 
  R.~Friedberg, T.~D.~Lee and Y.~Pang,
``Scalar Soliton Stars and Black Holes,''
  Phys.\ Rev.\ D {\bf 35}, 3658 (1987).


\bibitem{Gutowski:2004yv} 
  J.~B.~Gutowski and H.~S.~Reall,
  ``General supersymmetric AdS(5) black holes,''
  JHEP {\bf 0404}, 048 (2004)
  [hep-th/0401129].

  
\bibitem{Giusto:2012gt} 
  S.~Giusto and R.~Russo,
``Adding new hair to the 3-charge black ring,''
  Class.\ Quant.\ Grav.\  {\bf 29}, 085006 (2012)
  [arXiv:1201.2585 [hep-th]].
  
\bibitem{Vasilakis:2012zg} 
  O.~Vasilakis,
``Bubbling the Newly Grown Black Ring Hair,''
  JHEP {\bf 1205}, 033 (2012)
  [arXiv:1202.1819 [hep-th]].
  
  
\bibitem{Elvang:2005sa} 
  H.~Elvang, R.~Emparan, D.~Mateos and H.~S.~Reall,
  ``Supersymmetric 4-D rotating black holes from 5-D black rings,''
  JHEP {\bf 0508}, 042 (2005)
  [hep-th/0504125].
  
\bibitem{Gaiotto:2005xt} 
  D.~Gaiotto, A.~Strominger and X.~Yin,
  ``5D black rings and 4D black holes,''
  JHEP {\bf 0602}, 023 (2006)
  [hep-th/0504126].
  
\bibitem{Bena:2005ni} 
  I.~Bena, P.~Kraus and N.~P.~Warner,
  ``Black rings in Taub-NUT,''
  Phys.\ Rev.\ D {\bf 72}, 084019 (2005)
  [hep-th/0504142].
  
\bibitem{Bena:2004wv} 
  I.~Bena,
  ``Splitting hairs of the three charge black hole,''
  Phys.\ Rev.\ D {\bf 70}, 105018 (2004)
  [hep-th/0404073].
  
  \bibitem{Page:1979ga}
  D.~N.~Page,
 ``Green's Functions for Gravitational Multi - Instantons,''
  Phys.\ Lett.\ B {\bf 85} (1979) 369.
  
  \bibitem{Bena:2010gg}
  I.~Bena, N.~Bobev, S.~Giusto, C.~Ruef and N.~P.~Warner,
  ``An Infinite-Dimensional Family of Black-Hole Microstate Geometries,''
  JHEP {\bf 1103} (2011) 022
   [Erratum-ibid.\  {\bf 1104} (2011) 059]
  [arXiv:1006.3497 [hep-th]].
  
\bibitem{Mathur:2013nja} 
  S.~D.~Mathur and D.~Turton,
``Oscillating supertubes and neutral rotating black hole microstates,''
  JHEP {\bf 1404}, 072 (2014)
  [arXiv:1310.1354 [hep-th]].
  
\bibitem{Bena:2009ev} 
  I.~Bena, G.~Dall'Agata, S.~Giusto, C.~Ruef and N.~P.~Warner,
  ``Non-BPS Black Rings and Black Holes in Taub-NUT,''
  JHEP {\bf 0906}, 015 (2009)
  [arXiv:0902.4526 [hep-th]].
  
\bibitem{Bossard:2012ge} 
  G.~Bossard,
  ``Octonionic black holes,''
  JHEP {\bf 1205}, 113 (2012)
  [arXiv:1203.0530 [hep-th]].
  
\bibitem{Elvang:2004xi}
  H.~Elvang, R.~Emparan and P.~Figueras,
  ``Non-supersymmetric black rings as thermally excited supertubes,''
  JHEP {\bf 0502} (2005) 031
  [hep-th/0412130].
  
\end{thebibliography}
\end{document}